\documentclass[aps,prl,twocolumn,showpacs,10pt,superscriptaddress,preprintnumbers,nofootinbib]{revtex4-2}
\usepackage{amsmath}
\usepackage{physics, bm}
\usepackage{graphicx}
\usepackage{amssymb}
\usepackage[colorlinks,citecolor=red]{hyperref}

\usepackage{color}    
\usepackage{ulem}
\usepackage{multirow}

\newcommand{\beq}{\begin{equation}}
\newcommand{\eeq}{\end{equation}}
\newcommand{\bea}{\begin{eqnarray}}
\newcommand{\eea}{\end{eqnarray}}
\newcommand{\nn}{\nonumber}

\newcommand{\M}{\mathcal{M}}
\newcommand{\A}{\mathcal{A}}
\newcommand{\GeV}{{\rm GeV}}

\begin{document}

\preprint{
	{\vbox {			
		\hbox{\bf MSUHEP-23-019}
}}}
\vspace*{0.2cm}

\title{Single Transverse Spin Asymmetry as a New Probe of SMEFT Dipole Operators}

\author{Xin-Kai Wen}
\email{xinkaiwen@pku.edu.cn}
\affiliation{School of Physics, Peking University, Beijing 100871, China}

\author{Bin Yan}
\email{yanbin@ihep.ac.cn (corresponding author)}
\affiliation{Institute of High Energy Physics, Chinese Academy of Sciences, Beijing 100049, China}

\author{Zhite Yu}
\email{yuzhite@msu.edu}
\affiliation{Department of Physics and Astronomy,
Michigan State University, East Lansing, Michigan 48824, USA}

\author{C.-P. Yuan}
\email{yuanch@msu.edu}
\affiliation{Department of Physics and Astronomy,
Michigan State University, East Lansing, Michigan 48824, USA}

\begin{abstract}
Electroweak dipole operators in the Standard Model Effective Field Theory (SMEFT) are important indirect probes of
quantum effects of new physics beyond the Standard Model (SM),
yet they remain poorly constrained by current experimental analyses 
for lack of interference with the SM amplitudes in constructing cross section observables. 
In this Letter, we point out that dipole operators flip fermion helicities so are ideally studied through single transverse spin asymmetries.
We illustrate this at a future electron-positron collider with transversely polarized beams, where such effect exhibits as azimuthal 
$\cos\phi$ and $\sin\phi$ distributions which originate from the interference of the electron dipole
operators with the SM and are linearly dependent on their Wilson coefficients.
This new method can improve the current constraints on the electron dipole couplings by one to two orders of magnitude,
without depending on other new physics operators,
and can also simultaneously constrain both their real and imaginary parts, 
offering a new opportunity for probing potential $CP$-violating effects.
\end{abstract}

\maketitle

\noindent {\bf Introduction.}
The SMEFT provides a powerful systematic bottom-up approach in particle physics to 
parameterize possible new physics (NP) effects beyond the SM~\cite{Buchmuller:1985jz,Grzadkowski:2010es}. 
Measuring the corresponding Wilson coefficients to reveal hidden NP effects in the SM background 
has formed a major task of current and future colliders. 
Global SMEFT analyses with available data at the Large Hadron Collider (LHC) and other facilities have been carried out for various
subsets of the dimension-6 operators; see, {\it e.g.,} Refs.~\cite{Englert:2014uua,Falkowski:2015fla,Corbett:2015ksa,Cao:2015doa,Cao:2015qta,Cao:2015oaa,Cao:2016zob,Cirigliano:2016nyn,Alioli:2018ljm,Durieux:2018tev,Degrande:2018fog,Vryonidou:2018eyv,Durieux:2018ggn,Cao:2018cms,DeBlas:2019qco,Brivio:2019ius,Hartland:2019bjb,Du:2020dwr,Alioli:2020kez,Cirigliano:2021img,Ethier:2021bye,Miralles:2021dyw,Yan:2021tmw,Boughezal:2021tih,Cirigliano:2021peb,Cao:2021wcc,Du:2021rdg,Liao:2021qfj,Liu:2022vgo,deBlas:2022ofj,Dawson:2022bxd,Greljo:2022cah,Grunwald:2023nli}. 
Among them the electroweak dipole operators of light fermions, which is likely to entail important information about heavy particle interactions, are poorly constrained experimentally~\cite{Escribano:1993xr,daSilvaAlmeida:2019cbr,Boughezal:2021tih,Cao:2021trr,Boughezal:2023ooo,Grunwald:2023nli}. 
Unlike the current interaction vertices in the SM, dipole operators flip fermion helicities, so their interference 
with the former are suppressed by the light fermion masses if one only considers unpolarized rate observables
as in some typical analyses
~\cite{Beneke:2014sba,Craig:2015wwr,Ge:2016zro,Khanpour:2017cfq,Chiu:2017yrx,Khanpour:2017cfq,Durieux:2017rsg,Barklow:2017suo,Li:2019evl,Rao:2019hsp,Yan:2021tmw}.
The leading contribution from such operators then starts at $\order{1/\Lambda^4}$, where $\Lambda$ is the scale of the NP. 
This makes dipole operators harder to probe at a nominal experimental energy reach, 
should $\Lambda$ be far from the electroweak scale.

To address this issue, in this Letter we propose a novel approach to probing dimension-6 light fermion dipole operators 
via single transverse spin asymmetries (SSAs). 
We shall demonstrate it with the measurement of electron dipole operators at a transversely polarized electron-positron collider,
through the production processes of $Zh$, $Z\gamma$, $W^+W^-$, and $\mu^+\mu^-$ 
at a center-of-mass (c.m.) energy $\sqrt{s} = 250~\GeV$ for a future Higgs factory.
Even though the dipole operators yield different helicity amplitude structures from the SM so do not interfere with
the latter in constructing unpolarized cross sections, the transverse beam spin effects introduce exactly 
the interference of opposite helicity states, so are directly probing the interference of the dipole and SM operators. 
This results in nontrivial azimuthal $\cos\phi$ and $\sin\phi$ distributions that are linearly dependent on the 
Wilson coefficients of the dipole operators and serve as useful probes at $\order{1/\Lambda^2}$.
While such SSA observables are well known in the study of hadronic physics~\cite{Boer:2003xz} 
and heavy fermion productions~\cite{Kane:1978nd, Kane:1991bg, Yuan:1991nt, Dharmaratna:1996xd}, 
it is the first time for them to be used to enhance the study of dipole operators at future electron-positron colliders.

\vspace{3mm}
\noindent {\bf Transverse spin and azimuthal distributions.}
In the c.m. frame, the electron (positron) beam is chosen to be along the $\hat{z}$ ($-\hat{z}$) axis. 
The spin-dependent amplitude square is
\beq\label{eq:polarized-M2}
	\Sigma(\phi,\bm{s},\bar{\bm{s}})
	= 	\rho_{\alpha_1\alpha_1^\prime}(\bm{s})
		\rho_{\alpha_2\alpha_2^\prime}(\bar{\bm{s}})
		\M_{\alpha_1\alpha_2}(\phi)
		\M_{\alpha_1^\prime\alpha_2^\prime}^*(\phi),
\eeq
where the dependence on other kinematic variables has been suppressed.
Here $\alpha$'s denote beam helicities,
$\M_{\alpha_1\alpha_2}(\phi)$ is the helicity amplitude of $e^-_{\alpha_1} e^+_{\alpha_2}$ scattering into
the final state $Zh$, $Z\gamma$, $W^+W^-$, or $\mu^+\mu^-$ with a characteristic angle $\phi$, which is taken as the azimuthal angle of $Z$, $W^{\pm}$, or $\mu^{\pm}$ in the $e^+e^-$ c.m.~frame,
and we sum over repeated indices and final state spins.
The $\rho(\bm{s}) = (1 + \bm{s} \cdot \bm{\sigma})/2$ is the fermion spin density matrix with $\bm{\sigma}$ the Pauli matrices.
This defines the spin vector $\bm{s}$ ($\bar{\bm{s}}$) for the electron (positron) beam.
With respect to the same $\hat{x}$ and $\hat{y}$ axes, we parametrize them as
$\bm{s} = (\bm{s}_T, 0) = (b_T\cos\phi_0, b_T\sin\phi_0, 0)$ and 
$\bm{\bar{s}} = (\bar{\bm{s}}_T, 0) = (\bar{b}_T\cos\bar{\phi}_0, -\bar{b}_T\sin\bar{\phi}_0, 0)$,
where $b_T$~$(\bar{b}_T) > 0$ is the magnitude of the electron's (positron's) transverse spin and 
$\phi_0$ $(\bar{\phi}_0)$ its azimuthal orientation.
We have neglected effects from nonzero helicity polarization. 
Similar azimuthal distributions can arise from the interplay of a longitudinally polarized beam and a transversely polarized beam;
we leave that study for future.

Since $b_T$ and $\bar{b}_T$ enter Eq.~\eqref{eq:polarized-M2} in a linear way through the density matrices, 
$\Sigma(\phi,\bm{s},\bar{\bm{s}})$ can be decomposed as,
\beq
	\Sigma(\phi,\bm{s},\bar{\bm{s}})=\Sigma_{UU} + b_T\Sigma_{TU}(\phi) + \bar{b}_T\Sigma_{UT}(\phi) + b_T \bar{b}_T\Sigma_{TT}(\phi).
\label{eq:mm-phi}
\eeq
The $\Sigma_{UU}$ does not depend on the beam polarization nor the azimuthal angle $\phi$,
while $\Sigma_{TU}$ and $\Sigma_{UT}$ depend singly on the transverse spin of the electron and positron beams, respectively, 
$\Sigma_{TT}$ captures the transverse beam spin correlation, and they contain nontrivial $\phi$ dependence.
By using $\M_{\alpha_1\alpha_2}(\phi)=e^{i(\alpha_1-\alpha_2)\phi} \A_{\alpha_1\alpha_2}$, with $\A_{\alpha_1\alpha_2}$ being independent of $\phi$, 
the azimuthal dependence of $\Sigma_{TU, UT, TT}$ can be obtained from Eq.~\eqref{eq:polarized-M2} as~\cite{Hikasa:1985qi,Moortgat-Pick:2005jsx},
\begin{align}
	\Sigma_{TU}& 
		= \frac{1}{2} {\rm Re} \left[ e^{i(\phi - \phi_0)} \left(\A_{+-} \A_{--}^* + \A_{++} \A_{-+}^* \right)\right],\nn\\
	\Sigma_{UT}& 
		= \frac{1}{2} {\rm Re} \left[ e^{i(\phi - \bar{\phi}_0)} \left(\A_{+-} \A_{++}^* + \A_{--} \A_{-+}^*\right) \right],\nn\\
	\Sigma_{TT}& 
		= \frac{1}{2} {\rm Re} \left[ e^{-i(\phi_0 - \bar{\phi}_0)} \A_{++} \A_{--}^* \right.		\nn\\
		&\hspace{7em}	+ \left. e^{i (2\phi - \phi_0 - \bar{\phi}_0)} \A_{+-} \A_{-+}^* \right],
\label{eq:Amp}
\end{align}
where the subscript $\pm$ refers to $\alpha = \pm 1/2$.
So $\Sigma_{TU}$ and $\Sigma_{UT}$ contain $\cos\phi$ and $\sin\phi$ distributions, whereas
$\Sigma_{TT}$ gives $\cos2\phi$ and $\sin2\phi$ distributions.

Eq.~\eqref{eq:Amp} entails the double roles of the transverse spins $b_T$ and $\bar{b}_T$.
First they break the rotational invariance around the beam axis and induce nontrivial azimuthal distributions.
Second, the single spin observables $\Sigma_{TU}$ and $\Sigma_{UT}$ arise from the interference of 
amplitudes that differ by one single electron helicity flip. In the SM, this can happen for the gauge interactions only
via a mass insertion and is then suppressed by the electron mass. Within the SMEFT, in contrast, a helicity flip 
can be induced by dipole operators of electron without such mass suppression. 
Hence, the unique $\cos\phi$ and $\sin\phi$ distributions are linearly dependent on the corresponding Wilson coefficients.
The $\Sigma_{TT}$ part, on the other hand, is the interference of amplitudes that differ by double helicity flips, so can happen
for both the SM and SMEFT operators, depending quadratically on the dipole couplings.

\vspace{3mm}
\noindent {\bf SSAs from electron dipole operators.}
We consider the SMEFT dipole operator in this study,
\beq\label{eq:Leff}
	\mathcal{L}_{\rm eff} 
		= -\frac{1}{\sqrt{2}} \bar{\ell}_L \sigma^{\mu\nu}
			\left(g_1 \Gamma_B^e B_{\mu\nu}+g_2 \Gamma_W^e \sigma^a W_{\mu\nu}^a \right) \frac{H}{v^2} e_R
			+{\rm h.c.}
\eeq
where $\ell_L$ ($e_R$) is the first-generation left (right) handed lepton doublet (singlet) field
and $H$ is the Higgs doublet field, with $v=246~{\rm GeV}$ being the vacuum expectation value.
$B_{\mu\nu}$ and $W_{\mu\nu}^a$ are the gauge field strength tensors of the $U(1)_Y$ and $SU(2)$, respectively, 
with $g_1$ and $g_2$ the corresponding gauge coupling strengths. 
The dimensionless Wilson coefficients $\Gamma_B^e$ and $\Gamma_W^e$ quantify the dipole operator coupling strengths.

At tree level, the amplitudes $\A_{\alpha_1\alpha_2}$ in Eq.~\eqref{eq:Amp} are real up to possible complex phases of $\Gamma_B^e$ and $\Gamma_W^e$.
It is then evident that the real parts of $\Gamma_B^e$ and $\Gamma_W^e$ only generate $\cos(\phi - \phi_0)$ in $\Sigma_{TU}$ and
$\cos(\phi - \bar{\phi}_0)$ in $\Sigma_{UT}$. 
With nonzero imaginary parts,
they would break the $CP$ symmetry and also induce $\sin(\phi - \phi_0)$ and $\sin(\phi - \bar{\phi}_0)$.
Beyond tree level, quantum loop effects would also generate imaginary parts to the amplitudes $\A_{\alpha_1\alpha_2}$, which also contribute to the sine (cosine) modulations for real (imaginary) parts of the dipole couplings;
the magnitude is generally suppressed~\cite{Kane:1991bg} and will be neglected in this Letter. 
For convenience, in the following discussion, we define the electron dipole couplings to the photon and $Z$ boson,
$\Gamma_\gamma^e=\Gamma_W^e-\Gamma_B^e$ and
$\Gamma_Z^e=c_W^2\Gamma_W^e+s_W^2\Gamma_B^e$, respectively, where $s_W\equiv\sin\theta_W$ and $c_W\equiv\cos\theta_W$, with $\theta_W$ being the weak mixing angle.

Given the simple dependence on $\phi_0$ and $\bar{\phi}_0$ in Eq.~\eqref{eq:Amp}, it is sufficient to consider two experimental setups,
(1) aligned spin setup: $\phi_0 = \bar{\phi}_0 = 0$, and (2) opposite spin setup: $(\phi_0, \bar{\phi}_0) = (0, \pi)$.
Then, the azimuthal distribution of a two-body final state $i$ with transversely polarized lepton beams can be written as,
\begin{align}
	&\frac{2\pi \, d\sigma^i}{\sigma^i \, d\phi}
	= 1 + A_R^i(b_T, \bar{b}_T) \cos\phi + A_I^i(b_T, \bar{b}_T) \sin\phi 	\nn\\
	&\hspace{4.8em} + b_T \, \bar{b}_T \, B^i \cos2\phi
	+ \order{1/\Lambda^4},
\label{eq:dis}
\end{align}
where, as examples, $i = Zh$, $Z\gamma$, $W^+W^-$, or $\mu^+\mu^-$, and for simplicity, we have integrated over $\cos\theta$.
The $\phi$ angle is taken as the azimuthal angle of $Z/W^{\pm}/\mu^{\pm}$ in the $e^+e^-$ c.m. frame.
For the $Z\gamma$ production, we require the transverse momentum of the photon $p_T^\gamma>10~{\rm GeV}$ to avoid collinear singularity.
As indicated from Eqs.~\eqref{eq:mm-phi} and \eqref{eq:Amp}, the coefficients 
$A_{R,I}^i$ depend linearly on $b_T$, $\bar{b}_T$, and the dipole couplings $\Gamma_\gamma^e$ and $\Gamma_Z^e$,
with $A_R^i$ proportional to their real parts and $A_{I}^i$ their imaginary parts.
To the accuracy of $\order{1 / \Lambda^2}$, the coefficient $B^i$ has no dependence on the dipole couplings. 
It only receives contribution from the SM (and other non-dipole NP) interactions through the second term of $\Sigma_{TT}$ in Eq.~\eqref{eq:Amp}, 
while the first term vanishes in the massless lepton approximation.
The azimuthal $\cos\phi$ and $\sin\phi$ distributions
can therefore be used to give sensitive constraints on the dipole operators.

To extract the coefficients $A_R^i$ and $A_I^i$, it is convenient to define the SSA observables,
\begin{align}
	A_{LR}^i
		&\,= \frac{\sigma^i(\cos\phi>0) - \sigma^i(\cos\phi<0)}{\sigma^i(\cos\phi>0) + \sigma^i(\cos\phi<0)}
			= \frac{2}{\pi} A_R^i,\nn\\
	A_{UD}^i
		&\,= \frac{\sigma^i(\sin\phi>0) - \sigma^i(\sin\phi<0)}{\sigma^i(\sin\phi>0) + \sigma^i(\sin\phi<0)}
			= \frac{2}{\pi}A_I^i,
\label{eq:SSA}
\end{align}
which can be referred to as ``left-right'' and ``up-down'' asymmetries, respectively, for the event distribution in the transverse plane.
Here $\sigma^i(\cos\phi>0)$ is the integrated cross section with $\cos\phi>0$, etc.
By this construction, the coefficient $B^i$, which can receive contribution from the SM and other non-dipole NP operators,
does not contribute to the SSAs in Eq.~\eqref{eq:SSA}, so that this method is exclusively probing the electron dipole operators. 
By combining the measurements of $A_{LR}$ and $A_{UD}$, we can simultaneously determine the 
real and imaginary parts of the dipole couplings, allowing to also probe $CP$-violating effects.

An interesting constraint follows from properties of the $CP$ transformation,
under which the initial state $e^+$ and $e^-$ only exchanges their spins, 
the final states $W^+W^-$ and $\mu^+\mu^-$ are left invariant, whereas
$Zh$ and $Z\gamma$ flip their momentum directions. As a result,
\beq
	A^i(\bm{s}_T, \bar{\bm{s}}_T; \Gamma^e_{Z, \gamma})
	 = \pm A^i(\bar{\bm{s}}_T, \bm{s}_T; \Gamma^{e *}_{Z, \gamma}),
\eeq
which holds for both $A_R^i$ and $A_I^i$,
with ``$+$'' for $i = W^+W^-$ or $\mu^+\mu^-$, and ``$-$'' for $i = Zh$ or $Z\gamma$.
This constrains the coefficients $A_R^i$ and $A_I^i$ to have definite dependence on the beam spins,
namely,
\begin{align}
	(A_R^{WW, \, \mu\mu}, \, A_I^{Zh, \, Z\gamma}) & \, \propto \bm{s}_T + \bar{\bm{s}}_T, \nn\\
	(A_I^{WW, \, \mu\mu}, \, A_R^{Zh, \, Z\gamma}) &\, \propto \bm{s}_T - \bar{\bm{s}}_T.
\label{eq:A-spin-depend}
\end{align}
Therefore, we can take advantage of both beams being polarized to enhance the signals of SSAs.
For probing $A_R^{WW, \, \mu\mu}$ and $A_I^{Zh, \, Z\gamma}$ we shall use the aligned spin setup,
while $A_I^{WW, \, \mu\mu}$ and $A_R^{Zh, \, Z\gamma}$ are better measured with the opposite spin setup.

\vspace{3mm}
\noindent{\bf Numerical results and discussion.}
Now we present the projected constraints of the SSAs in Eq.~\eqref{eq:SSA} on the dipole couplings $\Gamma_{Z, \gamma}^e$ at an $e^+e^-$ collider. 
Since each final-state particle is color neutral, the parton showering and hadronization effects in their hadronic decay channels will not be 
correlated with the overall azimuthal distributions so will not affect the SSAs, nor will the detector effects, as we have verified explicitly.

The $Zh$ and $Z\gamma$ events at lepton colliders can be identified by the recoil mass method. 
We combine all decay modes of the $Z$-boson in the $Z\gamma$ channel, 
and for the $Zh$ process, we include all decay modes of the Higgs boson and 
$Z\to \ell^+\ell^-/jj$ with $\ell=e$, $\mu$, and $\tau$. 
To measure the $\phi$ distribution in the $W^+W^-$ production, we select the decay channel $W^+W^-\to jj\ell^{\pm}\nu_{\ell}(\bar{\nu}_\ell)$. 
The kinematic cuts for the decay products of final states can change the total event numbers, but shall not significantly alter the 
azimuthal angle distribution and will be neglected in the following analysis. 
The statistical uncertainties of the SSAs in Eq.~\eqref{eq:SSA} are given by
\beq
	\delta A_{LR, UD}^i 
		=\sqrt{\frac{1 - (A_{LR, UD}^i)^2}{N^i}}
		\simeq \frac{1}{\sqrt{N^i}},
\eeq
where $N^i$ is the number of the selected events of the final state $i$ for a certain spin setup after the kinematic cuts, and the ``$\simeq$'' in the second step 
takes the approximation $A_{LR, UD}^i  \simeq 0$ in the SM. 
We assume the systematic uncertainties are cancelled in the SSAs in this study~\cite{ILC:2013jhg}.

\begin{figure}
\centering
	\includegraphics[scale=0.33]{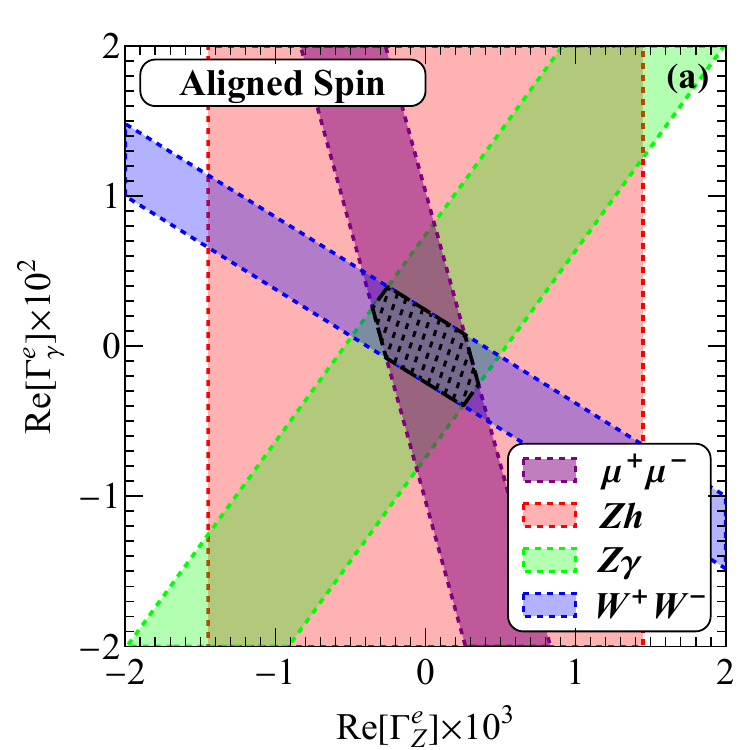}
	\includegraphics[scale=0.33]{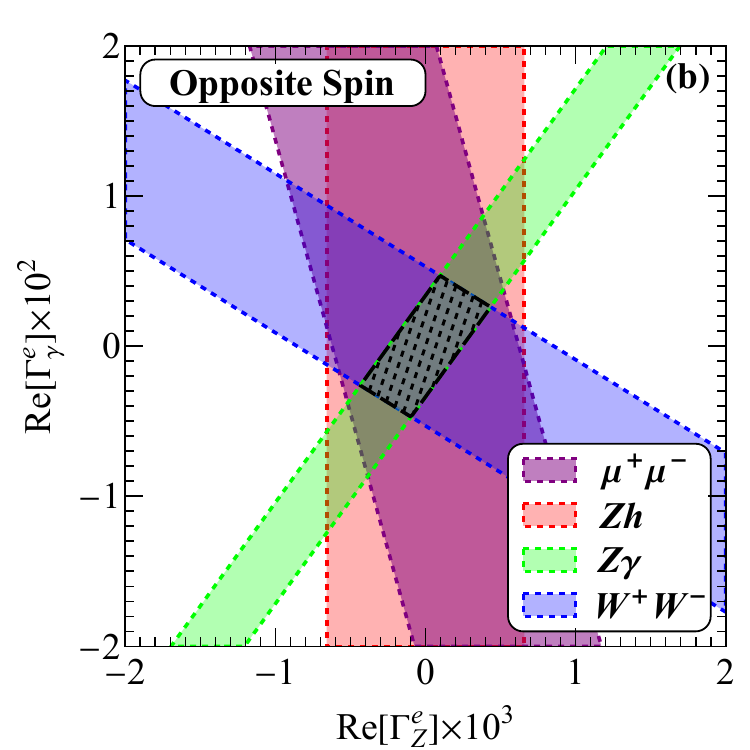}
	\includegraphics[scale=0.33]{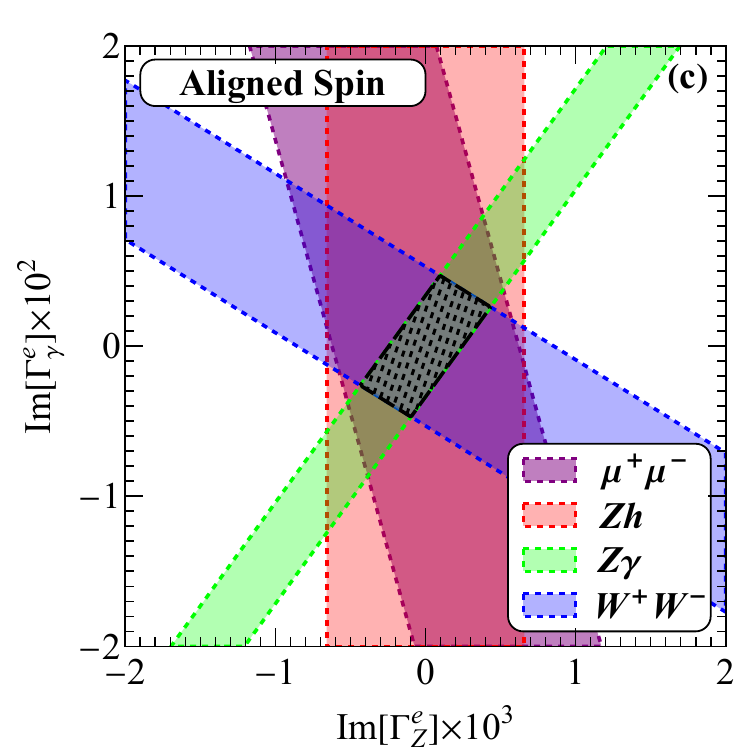}
	\includegraphics[scale=0.33]{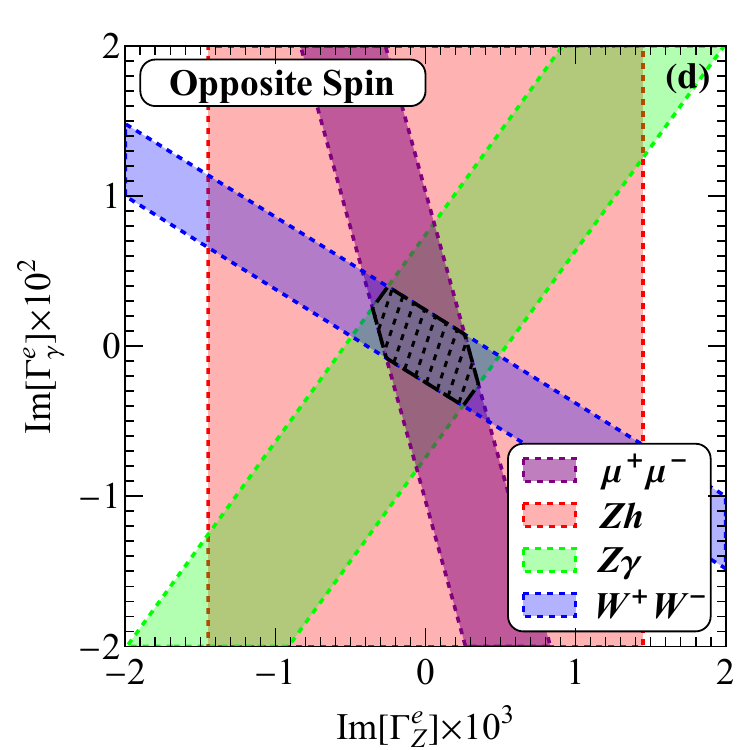}
\caption{Expected constraints on the real and imaginary parts of the electron dipole couplings $\Gamma_{Z,\gamma}^e$ from the SSAs $A_{LR, UD}$.
	The black bands are combinations of the four production channels.
}
\label{Fig:Limit}
\end{figure}

While there are several options in the market for a future lepton collider 
with a c.m.~energy $\sqrt{s}=240 \sim 250~{\rm GeV}$, 
such as the Circular Electron Positron Collider (CEPC)~\cite{CEPCStudyGroup:2018ghi}, 
the Future Circular Collider (FCC-ee)~\cite{FCC:2018evy}, 
and the International Linear Collider (ILC)~\cite{ILC:2013jhg}, 
none of them have explored the potential of using transversely polarized beams. 
As a benchmark study, we adopt the polarization parameters $(b_T, \, \bar{b}_T) = (0.8,\, 0.3)$ for our analysis~\cite{ILC:2013jhg}.

Figure~\ref{Fig:Limit} presents the expected constraints on the real and imaginary parts of the dipole couplings $\Gamma_{Z,\gamma}^e$ 
at 68\% confidence level (C.L.) for the $Zh$ (red bands), $Z\gamma$ (green bands), $W^+W^-$ (blue bands), and $\mu^+\mu^-$ (purple bands) channels 
with an integrated luminosity of $\mathcal{L}=5~{\rm ab}^{-1}$ at $\sqrt{s}=250~{\rm GeV}$ for each spin setup, 
under the SM assumption. 
The linear dependence of the SSAs in Eq.~\eqref{eq:SSA} on the dipole couplings lead to linearly shaped confidence regions,
among which the $Zh$ process is solely constraining $\Gamma_Z^e$, whereas 
the other channels are probing almost orthogonal linear combinations of 
$\Gamma_Z^e$ and $\Gamma_{\gamma}^e$.
As expected from Eq.~\eqref{eq:A-spin-depend}, the constraints of the aligned (opposite) spin setup on the real parts of 
$\Gamma_Z^e$ and $\Gamma_{\gamma}^e$
are the same as the opposite (aligned) spin setup on their imaginary parts.
It also confirms that the aligned (opposite) spin setup gives better constraints on $\Re\Gamma_{Z,\gamma}^e$ 
for the $W^+W^-$ and $\mu^+\mu^-$ ($Zh$ and $Z\gamma$) channels, and conversely for $\Im\Gamma_{Z,\gamma}^e$.

For the $Z\gamma$ and $\mu^+\mu^-$ processes, parity symmetry requires the dependence of $A_R$ ($A_I$) on $\Re\Gamma_{\gamma}^e$ ($\Im\Gamma_{\gamma}^e$) to be also proportional to 
the vector component of the SM $Ze\bar{e}$ coupling, $g_V^e=-1/2+2s_W^2\simeq -0.038$, which is nearly zero. 
As a result, the sensitivity to $\Gamma_Z^e$ is much stronger than $\Gamma_{\gamma}^e$. 
A similar conclusion holds in the $W^+W^-$ channel due to the larger $W W Z$ coupling than $W W \gamma$
and that the translation $\Gamma_W^e = \Gamma_Z^e + s_W^2 \Gamma_{\gamma}^e$ gives a heavier weight on $\Gamma_Z^e$.
After combining the four processes (black bands in Fig.~\ref{Fig:Limit}), 
we have the same constraints for the real and imaginary parts of the dipole couplings, 
with typical upper limits of order $\mathcal{O}(0.01\%)$ for $\Gamma_Z^e$ and $\mathcal{O}(0.1\%)$ for $\Gamma_{\gamma}^e$.  

Without the use of such SSAs, dipole operators only contribute to cross sections at $\mathcal{O}(1/\Lambda^4)$
and are also hard to be disentangled from other NP operators, which severely limits the ability to constrain their Wilson coefficients.
This is true for the studies in the literature using the Drell-Yan process at the LHC~\cite{Boughezal:2021tih} and $Z$-pole data at the LEP~\cite{Escribano:1993xr,ALEPH:2005ab},
which only constrain $|\Gamma_{Z,\gamma}^e|$ within $\mathcal{O}(1\%)$ when one operator is considered at a time.
They also lack the sensitivity to distinguish the real and imaginary parts of dipole couplings.
This makes the SSAs at the transversely polarized $e^+e^-$ collider a unique opportunity to constrain the dipole operators.

Before closing this section, we note that the real and imaginary parts of $\Gamma_{\gamma}^e$ 
can contribute to the electron's  anomalous magnetic (MDM) and electric dipole moment (EDM), respectively,
and have been severely constrained by the experimental measurements. 
It was found that $\Re \Gamma_{\gamma}^e \sim \mathcal{O}(10^{-6})$ and $\Im  \Gamma_{\gamma}^e \sim \mathcal{O}(10^{-13})$~\cite{Aebischer:2021uvt}, for arising from tree-level contributions.
It is also important to note that the photon and $Z$ dipole operators can mix under renormalization at the loop level. As a result, the measurements of electron MDM and EDM can provide strong constraints on the $Z$ dipole interactions when considering one operator at a time, which yields  $\Re\Gamma_Z^e\sim\mathcal{O}(10^{-2})$ and $\Im\Gamma_Z^e\sim\mathcal{O}(10^{-9})$~\cite{Alonso:2013hga,Kley:2021yhn,Aebischer:2021uvt}. However, the conclusions drawn from these constraints heavily depend on the theoretical assumptions made in the analysis. For instance, the presence of four-fermion operators can also contribute to the electron MDM and EDM at the loop level,  which could potentially affect the accuracy and reliability of the extracted values from data. 
In contrast, the measurements of SSAs proposed in this article do not depend on other dimension-six SMEFT operators except the dipole operators listed in Eq.~\eqref{eq:Leff}. Furthermore, the expected limits for the parameter $\Re\Gamma_Z^e$, cf. Fig.~\ref{Fig:Limit}, would be two orders of magnitude stronger than the constraints derived from the low-energy measurements of the electron MDM.
Finally, we notice that the Yukawa type operator $H^{\dag} H \bar{\ell}_L H e_R / \Lambda^2$ could also contribute to the $W^+W^-$ process, whose effect, however, is found to be negligibly small, as compared to the effect of dipole operators.

\vspace{3mm}
\noindent{\bf Conclusions.}
In this Letter, we proposed a novel approach to sensitively probing new physics dipole operators 
via single transverse spin asymmetry observables, using as a concrete example the 
constraining power of such observables on the electron dipole operators at a transversely polarized electron-positron collider.
By inducing a single fermion helicity flip, the interference of such operators with the SM generates
unique $\cos\phi$ and $\sin\phi$ distributions that are linearly dependent on their Wilson coefficients, 
with no contribution from the SM and other new physics operators.
This gives much stronger sensitivity to the electron dipole operators than other approaches in 
Drell-Yan and $Z$-pole processes, at the LHC and LEP, respectively, by one to two orders of magnitude.
It also allows to simultaneously determine the real and imaginary parts of those couplings, 
giving the opportunity to directly study potential $CP$-violating effects.
Our approach can be readily applied to similar studies at a muon collider~\cite{AlAli:2021let, Accettura:2023ked} and 
forthcoming Electron-Ion Collider~\cite{AbdulKhalek:2021gbh} when beams are transversely polarized.
With measurements of final-state spins~\cite{Yu:2021zpe, Yu:2022kcj}, 
it can also be applied to the studies at unpolarized colliders such as the LHC.

\vspace{3mm}
\noindent{\bf Acknowledgments.}
Xin-Kai Wen is supported in part by the National Science Foundation of China under Grants No.11725520, No.11675002 and No.12235001. 
Bin Yan is supported by the IHEP under Contract No. E25153U1. 
Zhite Yu and C.-P. Yuan are supported by the U.S. National Science Foundation under Grant No. PHY-2013791 
and C.-P. Yuan is grateful for the support from the Wu-Ki Tung endowed chair in particle physics.

\bibliographystyle{apsrev}
\bibliography{reference}

\end{document}